\documentclass{article}  
\usepackage{bigsky2007}
\usepackage{graphicx}
\frompage{000} \topage{000}                                              
\def\rightmark{Heaviness of Heavy Quarkonia}
\def\leftmark{J.L. Nagle for the PHENIX Collaboration}
 
\title{Heaviness of Heavy Quarkonia in Heavy Ion Collisions} 
\authors{
{J.L. Nagle$^1$ for the PHENIX Collaboration %
}\\[2.812mm]
{\normalsize
\hspace*{-8pt}$^1$ University of Colorado, \\ 
Boulder, CO, USA\\[0.2ex] 
}}
 
\abstract{
High energy heavy ion collisions at the Relativistic Heavy Ion Collider (RHIC) produce 
a novel medium characterized by an initial energy density over an order of magnitude above
the expected phase transformation value and that then evolves as a nearly inviscid liquid.  
Probing the medium with auto-generated particles is a key methodology to quantitatively 
determine
the medium properties.  Pairs of heavy quarks are an excellent probe since their spatial 
separation to form various quarkonia states spans the relevant range of color screening lengths in the
medium.  In this proceedings, we describe results from the PHENIX experiment on $J/\psi$ production
and discuss initial physics implications of the measurements.
}

 
\begin{document}

\maketitle
\setcounter{page}{1}

\section{Introduction}\label{intro}

In the first six years of data taking at RHIC, a great deal of information about the created medium
has been compiled~\cite{whitepapers,annualreview}.  
The ``near-perfect'' liquid QCD medium discovered at RHIC has now led to our desire to
study its properties in detail.  Heavy quarkonia studies are of major import (``heaviness'') to this undertaking.

When the initial cold nuclei inter-penetrate, one has hard parton-parton interactions that form heavy
flavor ($c\overline{c}$ and $b\overline{b}$).  These pairs are then passed through by the other side of the cold nuclei,
which may break up the correlation between the pair.  Afterwords, the pair is inside a hot nuclear medium
which may significantly screen or modify their attractive interactions, thus suppressing future heavy quarkonia
formation.  There are many factors influencing the final rate of quarkonia yields, including the feed down between states via decay (for example $\chi_c \rightarrow J/\psi + \gamma$).  Thus, another definition of heaviness (``not easily digested'') may also apply.

\section{PHENIX Experiment}

\begin{figure}[htb]
\vspace*{+0cm}
                 \insertplot{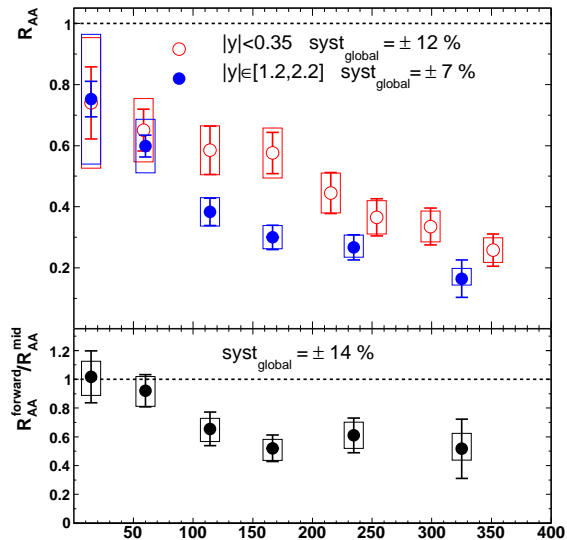}
\vspace*{-1cm}
\caption[]{
$J/\psi$ $R_{AA}$ versus the number of participating nucleons ($N_{part}$) for $Au+Au$ collisions.  Mid-rapidity and forward rapidity data points are shown as open and filled circles respectively.  In the lower panel is shown the rebinned ratio of the forward rapidity to mid-rapidity suppression values.
}
\label{fig_jpsi_raa}
\end{figure}

The PHENIX experiment at RHIC has measured the production of $J/\psi$ at mid-rapidity $|y|<0.35$ via their dielectron decay
and at forward rapidity $1.2 <|y|<2.2$ via their dimuon decay.  High statistics results are now available from
proton-proton reactions~\cite{phenix_pp_jpsi} and  results over the full
centrality range in $Au+Au$ reactions at $\sqrt{s_{NN}}=200$~GeV~\cite{phenix_auau_jpsi}.  It is notable that the signal
to background for the $J/\psi$ in central $Au+Au$ are 0.25 and 0.10 in the mid-rapidity and forward rapidity ranges, and
thus a careful accounting of systematic error contributions has been included.  The $Au+Au$ 
nuclear modification factors $R_{AA}$ are shown in Figure~\ref{fig_jpsi_raa}.


\section{Discussion of Results}

In many calculations of $J/\psi$ suppression in heavy ion reactions, a simple set of predictions arise.  If one assumes
the $J/\psi$ (or pre-cursor $c\overline{c}$ pair) are produced at rest in a static medium (with no time evolution), then
one can posit that if the local density (either $dE_{T}/dy$ or $dN_{ch}/dy$ for example) 
is greater than some threshold, then
no $J/\psi$ can be formed.  One can apply different effective thresholds for different quarkonia states, depending on their
binding energy in vacuum.  This picture leads to two predictions:  (1) a much larger suppression of quarkonia at 
RHIC energies compared with results from the lower $\sqrt{s_{NN}}$ at the fixed target CERN-SPS program, and 
(2) a larger suppression at mid-rapidity compared with forward rapidity, where the local density is lower.  

\begin{figure}[htb]
\vspace*{+0.0cm}
                 \insertplot{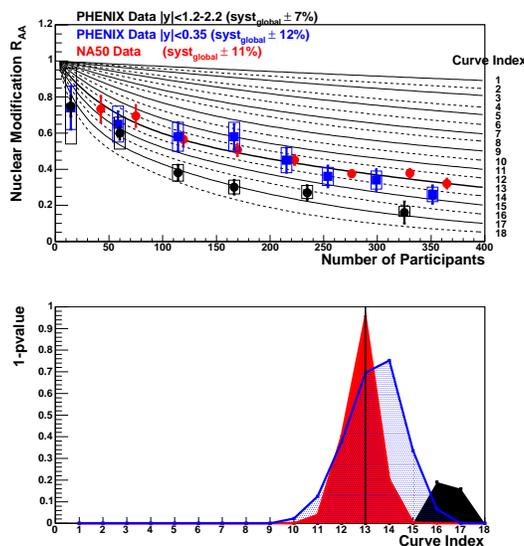}
\vspace*{-1cm}
\caption[]{Shown are the nuclear modification factor $R_{AA}$ as a function of collision centrality categorized
by the number of participating nucleons.  Data from the CERN-SPS from the NA50 experiment in $Pb+Pb$ reactions are shown
in addition to data from the PHENIX experiment in $Au+Au$ reactions at mid-rapidity and forward rapidity.  A series
of possible theoretical suppression curves is also shown, and then in the lower panel the $1-pvalue$ statistic between
each suppression curve and the experimental data are shown.
}
\label{fig_stat_test}
\end{figure}

As shown in Figure~\ref{fig_stat_test}, the results from the NA50 experiment~\cite{na50} at the CERN-SPS and the PHENIX results
at mid-rapidity at full RHIC energy are surprisingly compatible.  The statistical check of the p-value~\cite{PDG} for the different
theoretical curves (labeled by curve index) shows that both data sets are consistent within statistical
and systematic errors of having identical suppression patterns (and not just the central level).  
Note that in this most up-to-date data set from the NA50 experiment, there is no indication of non-monotonic derivatives
in the suppression pattern or a second ``drop'' for the most central events~\cite{na50_old}.  Thus, the real
feature is the overall smooth increase in suppression as one tends to more central collisions.
Also, shown are the 
PHENIX results at forward rapidity $1.2 < \eta < 2.2$, which indicate a significantly larger suppression level.  Both
of these observations from the experimental data are in contradiction with the simple predictions outlined above.

One need not be detered by these surprising findings, but rather consider alternate theoretical scenarios for
understanding the data and utilizing these probes.  One may consider two categories of theoretical explanations.
One is a simple explanation (``naturalness'') in which the common suppression pattern is a feature of a common
physics influence.  The second is an explanation that involves the fortuitous (or un-fortuitous) cancellation of
multiple effects (perhaps termed ``unnaturalness'').  Ocaam's Razor is paraphrased as ``all things being
equal the simplest solutions tends to be the best one.''  Of course, the answer does not have to be simple, but any 
explanation with cancellations needs to be tested rigorously with additional observations.

\begin{figure}[htb]
\vspace*{+0cm}
                 \insertplot{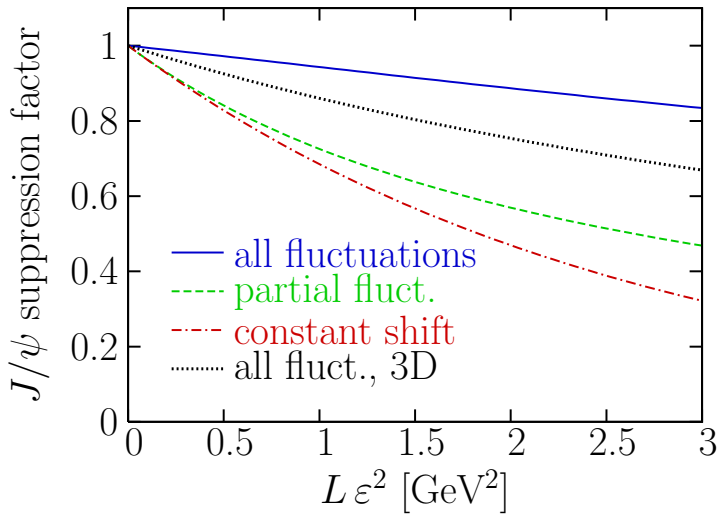}
\vspace*{-1cm}
\caption[]{
suppression of the total $J/\psi$ in nuclear collisions at
$\sqrt{s_{NN}}=200$ GeV as a function of 
$L\,\varepsilon^2$ for the constant shift (dashed-dotted), partial fluctuation 
(dashed) approximations, and the full 2D and 3D Gaussian random walk result (solid and dotted),
The full result gives much smaller suppression than the two approximations.
}
\label{fig_cold_supp}
\end{figure}

One naturalness category explanation is that the entire suppression effect is due to breakup of the $J/\psi$ 
pre-cursor $c\overline{c}$ state in the cold nuclear matter before the hot medium is even created.  It is 
essentially the same cold nucleus ($Au$ versus $Pb$) at the two energies (though moving with different
velocites relative to the $c\overline{c}$ pair).  The crossing time for cold nuclear matter is of order $1.0~fm/c$
at the CERN-SPS and $0.1~fm/c$ at RHIC.  In one such cold nuclear matter effect calculation~\cite{qiu}, 
they were able to
describe the entire NA50 suppression pattern for $Pb+Pb$ by assuming that the relative momentum $q^2$ between the 
charm and anticharm quark increases linearly with the cold nuclear matter path $L$ due to multiple scattering.
Then checking the overlap in momentum space of the $c\overline{c}$ with the quarkonia state yields a substantial
suppression, significantly larger than from extrapolating from proton-nucleus reactions.  It is notable that
this calculation assumes a coherent multiple scattering, where they assume that there is a constant increase
$\Delta q^2$ with distance $L$.  However, if this coherence assumption is relaxed or in the coherent case
there is not a simple shift in $q^2$, there is a substantial reduction in the suppression, as shown in 
Figure~\ref{fig_cold_supp}~\cite{glenn}.

\begin{figure}[htb]
\vspace*{+0.0cm}
                 \insertplot{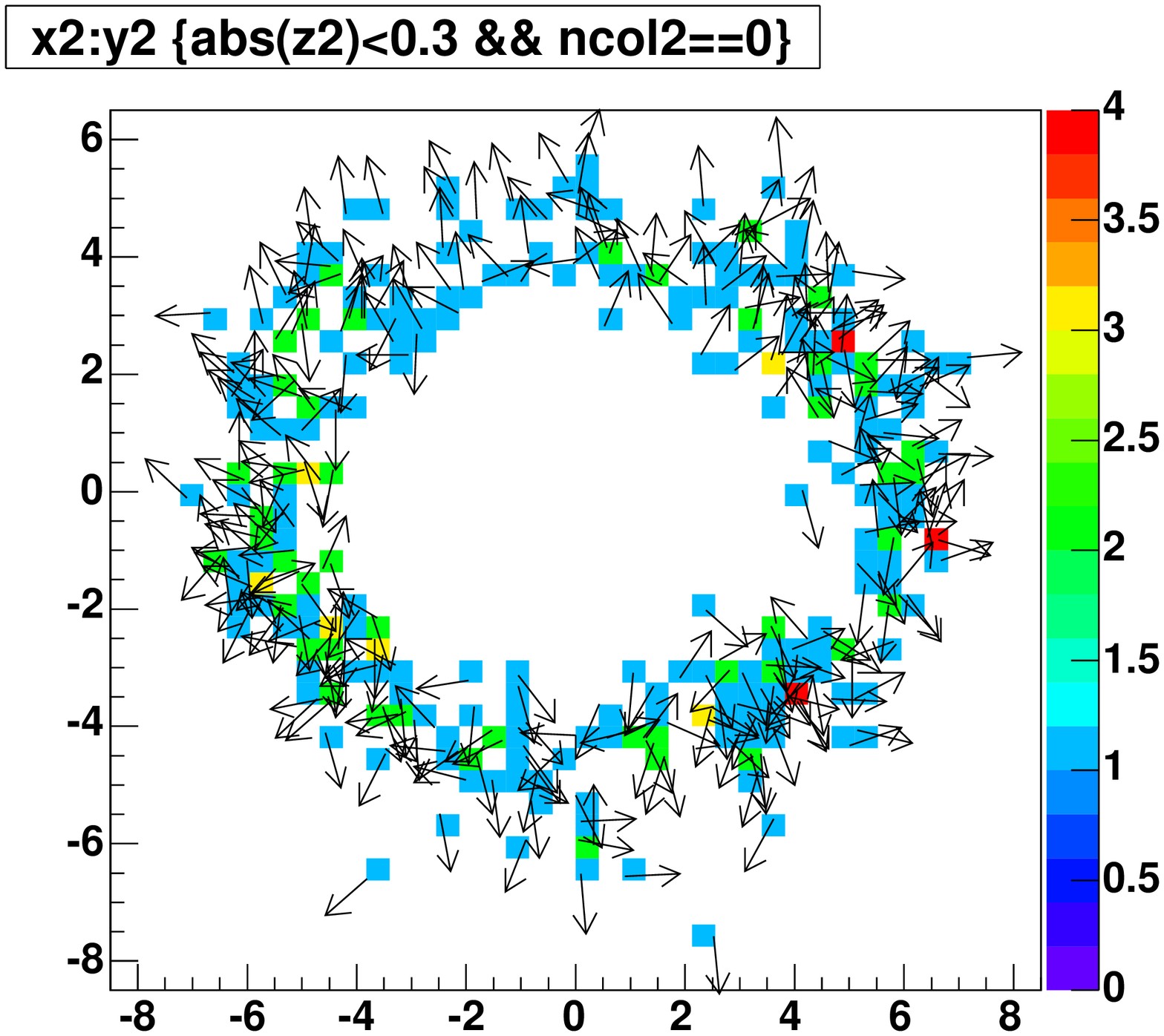}
\vspace*{-1cm}
\caption[]{
MPC simulation event for $Au+Au$ with impact parameter b=8 fm.  Shown is the t
ransverse plane spatial coordinates (x (fm) versus y(fm)) for charm quarks without any parton-parton scattering.
}
\label{fig_mpc}
\end{figure}

A second naturalness explanation is related to the large color opacity of the medium created at RHIC
(and perhaps also at the CERN-SPS).  In the case of a single quark or gluon propagating through the medium,
we have gauged the interaction with the strong color fields of the medium by the nuclear suppression $R_{AA}$ 
of $\pi^0$ for example.  It has been noted that the large color opacity leads to a bias of surface emmission
and thus a lack of sensitivity to the actually opacity value~\cite{lajoie}.  Perhaps the $c\overline{c}$ is propagating through
this opaque medium and emission is predominantly from the surface.  In this case, even though the 
opacity might be much larger for the medium created at RHIC, the suppression level would not be very different.  If
one assumed that the $c\overline{c}$ individual color charges are scattered separately (i.e. a complete
breakup of the initial quarkonia attraction), we can use the Molnar Parton Cascade (MPC)~\cite{molnar} to check for this effect.
Shown in Figure~\ref{fig_mpc} is the transverse plane spatial coordinates (x (fm) versus y(fm)) for charm quarks 
that suffered no parton-parton scattering.  One can see
a very large surface bias and also directional bias with the charm quarks preferentially moving outward.

\begin{figure}[htb]
\vspace*{+0.0cm}
                 \insertplot{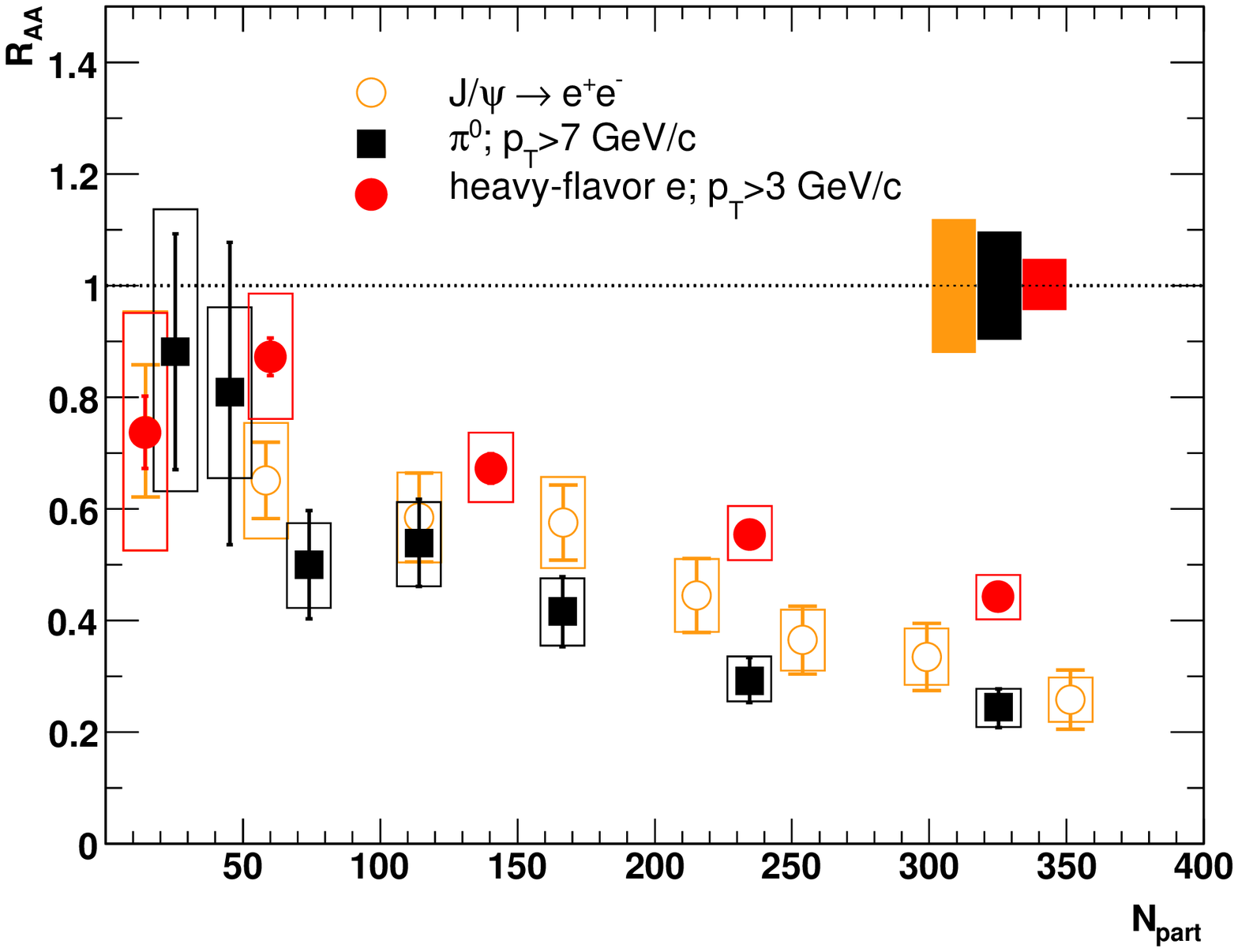}
\vspace*{-1cm}
\caption[]{Nuclear modification factor $R_{AA}$ versus $N_{part}$ for $J/\psi$ at midrapidity, $\pi^{0}$ 
at midrapidity, and non-photonic electrons with $p_{T}>$~3~GeV/c in $Au+Au$ reactions.
}
\label{fig_common_pattern}
\end{figure}

Another possible evidence of this scenario is the similar suppression patterns of $J/\psi$ at mid-rapidity, 
$\pi^{0}$ and heavy flavor (as measured by non-photonic electrons with $p_T>3$~GeV/c, as shown in 
Figure~\ref{fig_common_pattern}.

In principle, the impact of the hot nuclear medium needs to be separated from the cold nuclear
matter effects.  However, note that these physics impacts may not really factorize and may have
substantial correlations.  For example, a $c\overline{c}$ which has an increased $q^{2}$ from cold
matter interactions may then be more likely to dissociate in the hot matter afterwords.  Given
this caveat, some have noted that the cold nuclear matter effect can be parameterized by a $\sigma=4.18~mb$ 
$J/\psi$ breakup cross section~\cite{na50_cold}, while the RHIC data favor a value from $\sigma \approx 0-2~mb$~\cite{phenix_dau_jpsi}.
Thus, perhaps the hot nuclear matter suppression is larger at RHIC, but is masked by a smaller cold
nuclear matter effect.  Much higher statistics deuteron-gold results are needed at RHIC before making any
strong conclusion~\cite{raphael}.  However, if this were correct, the authors~\cite{kharzeev} speculate that if only
the $\psi'$ and $\chi_c$ states dissolve in medium (both at the CERN-SPS and RHIC), this might account for
the only slightly larger RHIC suppression (after correcting out the cold matter effects).  This might be
consistent with recent lattice QCD results indicating a much higher melting temperature for the $J/\psi$ than
previously thought.  Direct measurements of the $\chi_c$ are needed to test this.  In addition, it seems that
the suppression should turn on much more rapidly in peripheral events at RHIC, and this might be 
tested precisely with $J/\psi$ data from copper-copper reactions.

Another scenario involves the possibility of $J/\psi$ formation late in the time evolution via $c$ and $\overline{c}$ 
recombination or coalescence~\cite{recombination}.  
In this case, the $J/\psi$ that would have formed from originally produced
$c\overline{c}$ are much more suppressed at RHIC compared with the CERN-SPS, but this is almost exactly
compensated for by recombination as a new formation mechanism.  It is notable that this cancellation is completely
accidental and at intermediate energies (for example $\sqrt{s_{NN}}=62$~GeV) one might see larger suppression and
at higher energies (at the LHC) one might see no suppression at all (or even enhancement).  
An additional feature of these recombination $J/\psi$ is that they form from $c$ and $\overline{c}$ which are
typically within $\Delta y < 0.5$ and $\Delta p_{T} < 1$~GeV/c (close in momentum space).  Thus, these new
$J/\psi$ should contribute mostly at low $p_T$ and at mid-rapidity. Accordingly, an additional prediction is a narrowing
of the rapidity distribution (or a larger suppression at forward rapidity) and a steepening of the $p_T$ spectra.
No such significant $p_T$ modification is seen within errors, but there is a larger suppression at forward rapidity.
However, this prediction of rapidity narrowing depends on the original rapidity distribution of charm, which is
not yet constrained by any RHIC data.

Any recombination model must simultaneously match both the $J/\psi$ $p_T$ and $y$ distributions, but also
that of the open charm.  These models should predict a significant change in $J/\psi$ suppression for large $p_T >> 3~$GeV/c and also a large $J/\psi$ elliptic flow $v_2$~\cite{charmflow}.  Both of these should be testable with experimental data in the
next couple of years.  Another possible prediction from recombination is that states like the $\psi'$ which
have large suppression even at the CERN-SPS in peripheral $Pb+Pb$ collisions, might suddenly re-appear in
central $Au+Au$ at RHIC due to recombination.

The forward rapidity suppression may also have a separate explanation with a different type of cold nuclear
matter impact.  Perhaps due to shadowing of low-x partons in the heavy incoming nuclei, there may be a 
decrease in the initial $c\overline{c}$ production in $Au+Au$ relative to $p+p$.  If the density of these
low-x gluons is large enough, saturation physics may play a significant role.  However, such ideas needs to
be tested by checking consistently with future high statistics deuteron-gold collision data and with other
probes sensitive to similar low-x partons.

\section{Summary}

There are many exciting new results of heavy quarkonia from the PHENIX experiment that are of major import.
Full utilization of these quarkonia as a probe of the medium requires further theoretical understanding and
additional data to discriminate between competing explanations.  Measurements of multiple quarkonia states
and simultaneous matching of models to open and closed heavy flavor are critical.  Future measurements with
high statistics at RHIC II and at higher energies at the LHC should prove to be insightful.

\section*{Acknowledgments}
We thank the workshop organizers for providing an excellent atmosphere for physics
discussions.  We also acknowledge funding from the Division of Nuclear Physics
of the U.S. Department of Energy under Grant No. DE-FG02-00ER41152.


\vfill\eject
\end{document}